\documentclass[english,aps,manuscript]{revtex4}
\usepackage[T1]{fontenc}
\usepackage[latin1]{inputenc}
\usepackage{babel}

\makeatletter

\providecommand{\LyX}{L\kern-.1667em\lower.25em\hbox{Y}\kern-.125emX\@}

 \newenvironment{lyxlist}[1]
   {\begin{list}{}
     {\settowidth{\labelwidth}{#1}
      \setlength{\leftmargin}{\labelwidth}
      \addtolength{\leftmargin}{\labelsep}
      }}
   {\end{list}}

\newcommand{\bee}{\begin{equation}}
\newcommand{\ee}{\end{equation}}
\newcommand{\beea}{\begin{eqnarray}}
\newcommand{\eea}{\end{eqnarray}}

\makeatother
\begin{document}
{\centering \textbf{\Large On Potentials from Fluxes}\Large \par}

{\centering \vspace{0.3cm}\par}

{\centering {\large S. P. de Alwis\( ^{\dagger } \) }\large \par}

{\centering \vspace{0.3cm}\par}

{\centering Physics Department, University of Colorado, \\
 Boulder, CO 80309 USA\par}

{\centering \vspace{0.3cm}\par}

{\centering \textbf{Abstract}\par}

{\centering \vspace{0.3cm}\par}

{\centering We discuss the compactification of type IIB supergravity
with fluxes to generate a potential for the moduli. In particular
we resolve an apparent conflict with the no-go theorem for de Sitter
space. It is shown that a positive potential for certain moduli is
possible in situations where the volume modulus has no critical point.
We also point out that the derivation of the potential is strictly
valid only for a trivial warp factor. To go beyond that seems to require
the inclusion of all the Kaluza-Klein excitations. We end with a discussion
of the stabilization of the volume modulus.\par}

\vspace{0.3cm}


\vfill

\( ^{\dagger } \) {\small e-mail: dealwis@pizero.colorado.edu}{\small \par}

\eject

\section{Introduction}

There has been much activity since the mid eighties on compactifying
string theory on Ricci-flat manifolds in the presence of internal
fluxes. The main reason for this is that many of the moduli may be
stabilized by potentials generated by such fluxes. The first attempt
at moduli (actually the dilaton) stabilization using fluxes was the
paper of Dine et al \cite{Dine:1985rz}. More recently moduli potentials
from fluxes have been discussed in a large number of papers. (Some
recent papers are \cite{Lukas:1998tt}\cite{Lukas:1998yy}\cite{Taylor:1999ii}\cite{Kaloper:1999yr}\cite{Haack:1999zv}\cite{Curio:2000sc}\cite{Haack:2001jz}\cite{Dall'Agata:2001zh}\cite{Giddings:2001yu}\cite{DeWolfe:2002nn}).
Also it had been observed \cite{Becker:1996gj}\cite{Dasgupta:1999ss}\cite{Greene:2000gh}\cite{Grana:2000jj}\cite{Frey:2002hf}
that the conditions for obtaining supersymmetric solutions in (3 or
4 dimensional) space time required that some of the moduli of the
compactifications of M theory and F theory had to be fixed.

Perhaps the most interesting work in this direction is that of Giddings,
Kachru and Polchinski (GKP)\cite{Giddings:2001yu}. These authors
took type IIB string theory compactified on a Calabi-Yau manifold
in the presence of D3 branes and orientifold 3-planes. (GKP also considered
F-theory compactifications but we will not consider those here although
we expect the general conclusions to hold there too). They argued
that in this theory the potential is positive definite and is of the
no-scale type. They found that the dilaton may be stabilized at weak
coupling and that all the complex structure moduli would also be stabilized.
In effect, we seem to have a weak coupling stabilization of the dilaton
and other moduli, contrary to the expectation that moduli stabilization
requires string non-perturbative effects.

At first sight a positive potential would seem to be in conflict with
the no-go theorem for getting de Sitter space \cite{Gibbons:1984kp}
\cite{deWit:1987xg}\cite{Maldacena:2000mw}. To clarify the problem
we first look at the case of compactification with fluxes but without
local sources such as D-branes and orientifold planes. We argue that
the ansatz that goes into the proof of the no-go theorem is too restrictive.
With a more general ansatz (such generalizations have been also considered
independently in several papers, for example \cite{Cornalba:2002fi}\cite{Ohta:2003uw}
\cite{Townsend:2003fx}, in the context of cosmological solutions)
which allows the volume modulus to be time dependent, we see that
when there is no critical point for this modulus a positive potential
(for the other moduli) can be obtained.

Next we examine in detail the GKP proposal with D3 branes and orientifold
3-planes. Given the negative tension branes one can now have a conformally
Ricci flat internal space with non-zero flux. However within the static
ansatz for the metric that is used by GKP (and all other investigators
in the field) we show that there is no consistent way of getting the
potential. 

In order to compute the potential we put (4-space) time dependence
into the internal metric ansatz. The relevant equations are then corrected
by acceleration terms for the moduli (as well as velocity terms for
all moduli which can be set be set to zero at a given time by choice
of initial conditions). The essential point is that there is no critical
point for the volume modulus so it cannot be integrated out as is
(in effect) done in the discussion of the no-go theorem. One can then
recover the potential derived in \cite{Taylor:1999ii, Giddings:2001yu}.
However the ten dimensional equations of motion shows that this is
consistent only when the warp factor is constant and the combination
of fluxes which appears in the potential is constant. To remove this
restriction one needs to include Kaluza-Klein (KK) excitations in
the ansatz. We first look at the case of the KK excitations of the
volume modulus and encounter a puzzle. If one integrates out these
modes then one appears to get a negative definite potential! We argue
that this is due to the restricted nature of the KK ansatz and that
a resolution requires the inclusion of all the KK excitations. We
argue that consistency requires one to integrate out the complex structure
moduli and the dilaton along with the KK modes so that one is left
with supergravity coupled to the volume modulus with zero potential.
We conclude with some comments on attempts to stabilize all moduli.

\section{Compactification with fluxes and the no-go theorem}

Let us consider a D dimensional action for gravity coupled to a \( q=p+2 \)
form flux. (Our metric and curvature conventions are the same as in
GKP).

\begin{equation}
\label{Ddimaction}
S=\int d^{D}X\sqrt{g}\{R-\frac{1}{2.q!}F_{M_{1}...M_{q}}^{}F^{M_{1}...M_{q}}_{}\}.
\end{equation}

The flux \( F \) can be regarded as being dependent on some parameter
(field whose kinetic term we ignore). For instance in the type IIB
case this could be the combination of the RR and the NS 3-form fluxes
\( G_{3}=F_{3}-\tau H_{3} \) where \( \tau  \) is the axion-dilaton
field. 

The Einstein equation is \begin{equation}
\label{E0}
R_{MN}-\frac{1}{2}g_{MN}R=\frac{1}{2}T_{MN}.
\end{equation}

In trace-reversed form it is\begin{equation}
\label{E}
R_{MN}=\frac{1}{2}(T_{MN}-\frac{1}{8}g_{MN}T_{P}^{\, P}).
\end{equation}

The stress tensor is computed from 

\[
T_{MN}=-\frac{2}{\sqrt{|g|}}\frac{\delta S_{m}}{\delta g^{MN}}\]

where \( S_{m} \) is the action minus the curvature term. 

From the above action we have

\begin{equation}
\label{Eflux}
R_{MN}=\frac{1}{2q!}(qF_{MM_{2}...M_{q}}F_{N}^{\, .M_{2}...M_{q}}-\frac{1}{2}g_{MN}\frac{2q-2}{D-2}F^{2}).
\end{equation}

Now we wish to compactify the above theory on some \( d \) dimensional
compact manifold. We may do this by writing the metric in block diagonal
form up to a warp factor \( \Omega , \) 

i.e.\begin{equation}
\label{metric1}
ds^{2}=\Omega ^{2}(y)(\bar{g}_{\mu \nu }(x)dx^{\mu }dx^{\nu }+\bar{g}_{mn}(y)dy^{m}dy^{n}).
\end{equation}

In the above we've taken the coordinates on the compactification manifold
to be \( y^{m};m=1,...,d \) and the coordinates of the \( D-d \)
dimensional manifold to be \( x^{\mu };\mu =0,...,D-d-1. \)

Let us now take the projection of (\ref{Eflux}) onto the D-d manifold
using the metric (\ref{metric1}). We will also take the fluxes to
be non-zero only in the internal directions i.e. \( F_{\mu M_{2},,,M_{q}}=0 \),
\( F_{m_{1}...m_{q}}(y)\ne 0, \) assuming that \( q\le d \). Then
using standard formulas for conformal transformations we have,

\begin{equation}
\label{Ebar}
\bar{R}_{\mu \nu }(x)-\frac{1}{D-2}\Omega ^{-D+2}\bar{\nabla }^{2}_{y}\Omega ^{D-2}\bar{g}_{\mu \nu }(x)=-\frac{1}{2}\bar{g}_{\mu \nu }(x)\Omega (y)^{-2q+2}\frac{2q-2}{D-2}\frac{1}{2q!}\overline{F_{m_{1}...m_{q}}F^{m_{1...}m_{q}}}.
\end{equation}

Now the effective \( D-d \) dimensional trace reversed Einstein equation
for gravity (with metric \( \bar{g}) \) coupled to scalar fields
with a potential \( V \) (setting Kinetic terms to zero by taking
the fields to be constant) is\begin{equation}
\label{pot}
\bar{R}_{\mu \nu }(x)=\frac{\bar{g}_{\mu \nu }}{D-d-2}V
\end{equation}

Multiplying (\ref{Ebar}) by \( \Omega ^{D-2} \)(this is the factor
of Omega coming from the ten dimensional action) and integrating over
the compact internal manifold and comparing with the above we get
the effective potential,

\begin{equation}
\label{effpot}
V=-\frac{D-d-2}{2}\frac{2q-2}{D-2}\frac{1}{2q!}\int \sqrt{\bar{g}(y)}\Omega ^{D-2q}\overline{F_{m_{1}...m_{q}}F^{m_{1...}m_{q}}}/\int \sqrt{\bar{g}(y)}\Omega ^{D-2}.
\end{equation}
 As we will discuss in detail later, this is the potential for the
parameters such as \( \tau  \) and the internal components of the
metric provided that the determinant of the internal metric is fixed
at a critical point i.e. has ceased to move. Note that the denominator
in this expression has to be well-defined if a four dimensional reduction
is to exist. It is (up to numerical constants) essentially the four
dimensional Planck scale. An immediate consequence of this expression
is that if, as is the case for physically relevant cases, \( \frac{(D-d-2)(q-1)}{D-2} \)
is positive, the potential generated by internal fluxes is negative
definite.

Several points should be emphasized here. 

\begin{lyxlist}{00.00.0000}
\item [1.]The sign of the potential is opposite to what one might have expected
if one directly tried to reduce the D dimensional action to D-d dimensions
assuming an internal manifold of vanishing Ricci curvature. The point
is that the Ricci curvature in the D dimensional metric cannot be
assumed to be zero. That would contradict the projection to the internal
manifold of the D dimensional Einstein equation (\ref{Eflux}). The
internal flux gives positive Ricci curvature to the internal manifold
and so it can at best only be conformally Ricci flat.
\item [2.]The result does not depend on any assumptions about the \( \bar{g}_{mn}(y) \)
metric, for instance that it is (conformally) Ricci flat. The point
is that \( \Omega  \) in the above expression needs to be determined
from the internal space projection of the D dimensional Einstein equation.
Thus one needs some assumption about the geometry of the internal
space. We will discuss later the conditions under which one can assume
that \( \bar{g} \) or a conformally related metric is Ricci flat. 
\item [3.]Observation 1. is not new. In one form or the other it has been
known since the days of the earliest flux induced compactifications
such as that of Freund and Rubin \cite{Freund:1980xh} ( see also
\cite{Duff:1986hr}). It is in fact an aspect of the so-called no-go
theorem for getting de Sitter space from compactified higher dimensional
supergravity first derived in \cite{Gibbons:1984kp} (see also \cite{deWit:1987xg}\cite{Maldacena:2000mw}).
In other words what we seem to have shown is that if the higher dimensional
theory obeys the strong energy condition (\( R_{00}\geq 0 \) in every
frame) then the compactified theory also obeys this condition.
\end{lyxlist}
It is worth elaborating on point 2 above since there are some potentially
confusing issues associated with it. For concreteness, and also because
it is the case that is relevant to the string theory related discussions
of many papers involving flux compactifications in the literature,
we will specialize to the case of \( D=10,\, d=6 \). 

As mentioned in point 2 above, the internal manifold can only be conformally
Ricci flat at best. First we ask whether the internal metric \( \bar{g}_{mn} \)
can be made Ricci flat, i.e. \( \bar{R}_{mn}=0 \). If so we have
from the internal space projection of (\ref{Eflux}) (traced over
the internal space) 

\[
R^{(6)}=-40e^{-2\omega }\overline{\partial _{p}\omega \partial ^{p}\omega }-14e^{-2\omega }\bar{\nabla }^{2}\omega =\frac{q+3}{8q!}F_{q}^{2}\, \]
where \( e^{2\omega }=\Omega . \) Let us integrate this over the
internal manifold after multiplying by \( e^{2\omega }. \) Then the
second term (on the left hand side of the second equality) vanishes
because the manifold is compact and we then have a contradiction since
the LHS is negative definite while the RHS is positive definite. The
only solution is zero flux and constant warp factor. Of course we
do not need the middle equation here to compute the effective action.
The point of giving it is to show that the assumption of Ricci-flatness
for the metric \( \bar{g}_{mn} \) is not compatible with non-zero
flux.

Substituting this and the expression for \( R^{(4)} \) in the barred
metric, in the ten dimensional action we get the effective action
in 4D,\[
S=\int \sqrt{\bar{g}^{(4)}}d^{4}x\left[ \left( \int \sqrt{\bar{g}^{(6)}}d^{6}ye^{8\omega }\right) \bar{R}^{(4)}(x)+\frac{q-1}{8q!}\int \sqrt{\bar{g}^{(6)}}e^{(10-2q)\omega }\overline{F_{q}^{2}}\right] .\]

So we have a potential in agreement with (\ref{effpot}) as of course
it should. The point of the exercise was to emphasize that the calculation
had nothing to do with the assumption of Ricci-flatness of the internal
metric which indeed is invalid in this case. 

To highlight the latter issue let us now require Ricci flatness not
in the metric \( \bar{g}_{mn} \) as above but in the metric \( \tilde{g}_{mn}=e^{4\omega }\bar{g}_{mn} \)
. In other words we now write the 10D metric as \begin{equation}
\label{metric2}
ds^{2}=g_{MN}dx^{M}dx^{N}=e^{2\omega (y)}\bar{g}_{\mu \nu }(x)+e^{-2\omega (y)}\tilde{g}_{mn}(y)dy^{m}dy^{n}.
\end{equation}

This is in fact the form of the used by GKP \cite{Giddings:2001yu}.
Let ask whether the assumed Ricci flatness of the tilde metric is
compatible with non-zero flux. The equation for the internal curvature
now becomes,\begin{equation}
\label{r6}
R^{(6)}=-8\partial _{m}\omega \partial ^{m}\omega +6e^{+2\omega }\tilde{\nabla }^{2}\omega =\frac{q+3}{8q!}F_{q}^{2}.
\end{equation}

where we have put \( \tilde{R}_{mn}=0. \) Let us now multiply the
above by \( e^{-2\omega } \) and integrate to get, \[
\int \sqrt{\tilde{g}^{(6)}}d^{6}ye^{-2\omega }R^{(6)}=-8\int \sqrt{\tilde{g}^{(6)}}d^{6}ye^{-2\omega }\partial _{m}\omega \partial ^{m}\omega =\frac{q+3}{8q!}\int \sqrt{\tilde{g}^{(6)}}d^{6}ye^{-2\omega }F_{q}^{2}.\]

Thus we see that if one imposes Ricci flatness on the metric \( \tilde{g}_{mn} \)
then this makes the expression for the integrated internal curvature
negative definite while the flux makes it positive definite. Thus
again consistency requires vanishing flux and constant warp factor.

In order to understand better what is going on we need to consider
a more general metric ansatz that allows us to take into account the
4-space-time dependence of the moduli fields. To this end we write\[
ds^{2}=g_{\mu \nu }(x,y)dx^{\mu }dx^{\nu }+g_{mn}(x,y)dy^{m}dy^{n}.\]

Computing the Riemann tensor for this metric and contracting we have
the following results:\begin{eqnarray}
R_{\mu \nu }= & R_{\mu \nu }^{(4)}-\frac{1}{4}g^{\pi \sigma }g^{rm}\partial _{r}g_{\pi \sigma }\partial _{m}g_{\mu \nu }+\frac{1}{2}g^{pm}g^{\rho \lambda }\partial _{m}g_{\rho \nu }\partial _{p}g_{\lambda \mu } & \nonumber \\
-\frac{1}{4}g^{pm}g^{rn}\partial _{\nu }g_{mr}\partial _{\mu }g_{np} & -\frac{1}{2}\nabla _{p}^{(6)}(g^{pm}\partial _{m}g_{\mu \nu })-\frac{1}{2}\nabla ^{(4)}_{\nu }(\partial _{\mu }\ln g^{(6)}) & \label{Rmunu} \\
g^{mn}R_{mn}= & R^{(6)}-\frac{1}{4}g^{\rho \mu }\partial _{\rho }\ln g^{(6)}\partial _{\mu }\ln g^{(6)} & \nonumber \\
-\frac{1}{4}g^{\pi \sigma }g^{\rho \lambda }g^{mn}\partial _{m}g_{\rho \sigma }\partial _{n}g_{\lambda \pi } & -\frac{1}{2}g^{\pi \sigma }\nabla _{\pi }\partial _{\sigma }\ln g^{(6)}-\frac{1}{2}g^{mn}\nabla _{n}^{(6)}\partial _{m}\ln g^{(4)}.\label{Rmn} 
\end{eqnarray}
 \begin{eqnarray}
R_{\mu n}=\frac{1}{2}g^{\pi \lambda }\nabla _{\pi }^{(4)}\partial _{n}g_{\lambda \mu }-\partial _{n}\partial _{\mu }\ln \sqrt{g^{(4)}} & +\frac{1}{4}g^{\pi \lambda }g^{rp}\partial _{r}g_{^{_{\pi \lambda }}}\partial _{\mu }g_{pn} & -\frac{1}{4}g^{\pi \sigma }g^{rp}\partial _{p}g_{\mu \pi }\partial _{\sigma }g_{rn}\nonumber \\
\frac{1}{2}\nabla _{p}^{(6)}g^{pm}\partial _{\mu }g_{mn}-\partial _{n}\partial _{\mu }\ln \sqrt{g^{(6)}} & +\frac{1}{2}\partial _{\rho }\ln \sqrt{g^{(6)}}g^{\rho \nu }\partial _{n}g_{\nu \sigma } & -\frac{1}{4}g^{pm}\partial _{\rho }g_{mn}g^{\rho \nu }\partial _{p}g_{\nu \mu }\label{Rmun} 
\end{eqnarray}

In the above the superscripts \( (4), \) (6) denote the fact that
the corresponding quantities are restricted respectively to the external
4-space-time and the internal 6-space. It is important to note that
the second time derivative of the internal metric volume factor (
but of no other internal metric component) appears in these two equations.
Hence we see that equation (\ref{r6}) is in fact a second order (in
time) differential equation for \( g^{6} \) - the determinant of
the internal metric, of the form,\[
-\frac{1}{2}g^{\pi \sigma }\nabla _{\pi }\partial _{\sigma }\ln g^{(6)}+...=-R^{(6)}+\frac{1}{4}g^{\pi \sigma }g^{\rho \lambda }g^{mn}\partial _{m}g_{\rho \sigma }\partial _{n}g_{\lambda \pi }+\frac{1}{2}g^{mn}\nabla _{n}^{(6)}\partial _{m}\ln g^{(4)}+\frac{q+3}{8q!}F_{q}^{2}.\]

The ellipses denote terms which are first order in 4 space-time derivatives.
The left hand side is then the force acting on the determinant of
the internal metric. Integrating over the internal manifold, we see
that if the (integrated) internal curvature scalar \( R^{6} \) is
positive then there is a critical point for the volume modulus of
the 6 dimensional space. This would be the case when for instance
the internal manifold is a sphere (as in the Freund-Rubin case). However
if the internal manifold is conformal to a compact Ricci flat manifold
then as in (\ref{r6}) the first three terms on the RHS of the above
equation are replaced by positive definite terms plus a term which
integrates to zero on a compact manifold (in the tilde metric) and
there is no critical point. Of course a sphere is conformally flat
but the corresponding flat space is non-compact Euclidean space and
so there is no contradiction. \emph{In fact the proof of the absence
of a critical point will not be valid for manifolds which have a metric
that is conformal to a non-compact Ricci-flat manifold.}

For future reference let us introduce some additional notation. First
we put \begin{equation}
\label{warp}
g_{mn}=e^{2\omega (y)-2u(x)}\tilde{g}_{mn}(x,y)
\end{equation}
 where we take the tilde metric to be such that \( \partial _{\mu }\tilde{g}^{(6)}=0. \)
So \( u(x) \) is the four dimensional field which controls the volume
of the internal space - the volume modulus. \( \omega (y) \) on the
other hand is a warp factor which is actually defined by writing the
four dimensional metric as \[
g_{\mu \nu }(x,y)=e^{2\omega (y)-6u(x)}\tilde{g}_{\mu \nu }(x).\]
The dependence on \( u \) in the above is determined by requiring
that this modulus does not mix with the 4D graviton. Thus our final
form for the metric is\begin{equation}
\label{metric3}
ds^{2}=e^{2\omega (y)-6u(x)}\tilde{g}_{\mu \nu }(x)dx^{\mu }dx^{\nu }+e^{-2\omega (y)+2u(x)}\tilde{g}_{mn}(x,y)dy^{m}dy^{n}
\end{equation}

with \( \partial _{\mu }\det \tilde{g}_{mn}=0. \) Then we may rewrite
(\ref{Rmunu},\ref{Rmn}) as follows:\begin{eqnarray}
R_{\mu \nu }= & \tilde{R}_{\mu \nu }^{(4)} & -e^{4\omega (y)-4u(x)}\tilde{\nabla }^{2}_{y}\omega (y)\tilde{g}_{\mu \nu }+3\tilde{\nabla }^{2}_{x}u(x)\tilde{g}_{\mu \nu }+...\label{Rmunu2} \\
g^{mn}R_{mn}= & e^{2\omega (y)-2u(x)}\tilde{R}^{(6)} & -8e^{2\omega (y)-2u(x)}\tilde{g}^{mn}\partial _{m}\omega \partial _{n}\omega +6e^{2\omega (y)-2u(x)}\tilde{\nabla }^{2}_{y}\omega (y)\nonumber \\
 &  & -6e^{-2\omega (y)+6u(x)}\tilde{\nabla }^{2}_{x}u(x)+...\label{Rmn2} 
\end{eqnarray}

Again the ellipses represent terms involving first order derivatives
in 4-space-time. It is important to note that the second time derivatives
of the volume modulus \( u(x) \) enters into the expression for (\ref{Rmunu})
and that no other second time derivative term of the internal metric
enters there. What this means is that the expression for the potential
obtained earlier by trace reversing the ten dimensional equation and
projecting the 4-space-time components make sense only when the volume
modulus is stabilized. This would be the case when we compactify on
a sphere for instance, but for an internal manifold that is conformal
to a \emph{compact} Ricci-flat manifold the volume modulus cannot
be stabilized. 

We also note for future reference that since \( R_{\mu \nu }\sim 3\tilde{\nabla }^{2}_{x}u(x)\tilde{g}_{\mu \nu }, \)
\( \tilde{g}^{\lambda \sigma }R_{\lambda \sigma }\tilde{g}_{\mu \nu }\sim 12\tilde{\nabla }^{2}_{x}u(x)\tilde{g}_{\mu \nu } \),
\( g^{mn}R_{mn}\tilde{g}_{\mu \nu }\sim -6\tilde{\nabla }^{2}_{x}u(x)\tilde{g}_{\mu \nu } \)
the 4-space-time projection of the Einstein equation in its original
form i.e.\begin{equation}
\label{E1}
R_{\mu \nu }-\frac{1}{2}g_{\mu \nu }g^{\lambda \sigma }R_{\lambda \sigma }=\frac{1}{2}T_{\mu \nu }+\frac{1}{2}g_{\mu \nu }g^{mn}R_{mn},
\end{equation}

has no second time derivatives of the moduli. Indeed the \( u \)
dependent conformal factor in \( g_{\mu \nu }(x,y) \) was designed
to achieve this. Thus one may read off the effective potential directly
from this equation as being the term independent of space-time derivatives
- after integrating over the internal manifold. What one finds for
the above case of compactification with fluxes is that if the manifold
has positive curvature in the original metric, then the potential
is of indefinite sign (the fluxes giving a positive definite term
and the internal curvature a negative definite term) and there is
a critical point for the volume modulus. Let us express the above
equation in terms of the metric ansatz (\ref{metric3}) \begin{eqnarray*}
\int d^{6}y\sqrt{\tilde{g}^{(6)}}e^{-4\omega (y)}(\tilde{R}_{\mu \nu }-\frac{1}{2}\tilde{g}_{\mu \nu }\tilde{g}^{\lambda \sigma }\tilde{R}_{\lambda \sigma })= & -\frac{1}{2}\tilde{g}_{\mu \nu }\int d^{(6)}ye^{-4\omega (y)} & \\
(\frac{1}{2q!}F_{m_{1}...m_{q}}F^{m_{1}...m_{q}}+8e^{2\omega -2u}\tilde{g}^{mn}\partial _{m}\omega \partial _{n}\omega  & -\tilde{R}^{(6)}e^{2\omega -2u}) & +...
\end{eqnarray*}

Where the ellipses denote kinetic terms. Clearly if \( \tilde{R}^{6)} \)is
negative or zero, the potential is positive definite, but now there
is no critical point for the volume modulus as can be seen clearly
from the equation for the internal curvature which now an equation
of motion for \( u \). \begin{eqnarray*}
6e^{2\omega (y)-2u(x)}\tilde{\nabla }^{2}_{y}\omega (y) & -6e^{-2\omega (y)+6u(x)}\tilde{\nabla }^{2}_{x}u(x)+... & =\frac{q+3}{8q!}e^{-2q(\omega -u)(}F_{m_{1}...m_{q}}\tilde{F}^{m_{1}...m_{q}}\\
 & +8e^{2\omega (y)-2u(x)}\tilde{g}^{mn}\partial _{m}\omega \partial _{n}\omega  & -e^{2\omega (y)-2u(x)}\tilde{R}^{(6)}
\end{eqnarray*}

So integrating over the internal manifold (after multiplying by \( e^{-2\omega } \)
) we see that if \( \tilde{R}^{(6)}\le 0 \), the left hand side cannot
vanish, meaning that there is no critical point for the potential
for \( u(x). \)

\section{Type IIB with fluxes and local sources}

Let us now look at the work of GKP \cite{Giddings:2001yu} where the
compactification of low energy type IIB string theory was considered.
In this work it was argued that in order to obtain a flat space solution
one needs to introduce negative tension sources, or introduce an F-theory
compactification with dilaton gradients and seven branes. This was
essential in order to evade the above mentioned no-go theorem \cite{Gibbons:1984kp}\cite{deWit:1987xg}\cite{Maldacena:2000mw}.
Nevertheless in calculating the potential the authors just dimensionally
reduced the flux term in the ten dimensional action and got a positive
definite potential. However as we've seen above, the effective potential
in 4 dimensions coming just from the fluxes is in fact negative definite
if the volume modulus can be stabilized. It is precisely this that
prevents one from getting 4 dimensional de Sitter space starting from
a higher dimensional (greater than 6D) action. On the other hand as
has been shown by several authors there exists a flat 4D supersymmetric
solution solution with fluxes and one might wonder whether these can
be obtained as the minimum of a \emph{four} dimensional potential
with the all Kaluza-Klein (KK) modes integrated out. In the following
we will discuss the condition under which this is possible along the
lines of the previous section.

Two caveats are in order here:

\begin{itemize}
\item We focus on the D3/O3 brane case of GKP and ignore the F-theory construction.
It is expected that this gives a similar result.
\item All our arguments ignore the effect of higher derivative terms - as
in GKP.
\end{itemize}
The type IIB action in Einstein frame is (with \( 2\kappa ^{2}_{10}=1 \)
) \begin{eqnarray*}
S & = & \int d^{10}X\sqrt{-g}\{R-\frac{1}{2\tau _{I}^{2}}\partial _{M}\tau \partial ^{M}\tau -\frac{1}{2.3!\tau _{I}}G_{MNP}G^{MNP}-\frac{1}{4.5!}\tilde{F}_{MNPQR}\tilde{F}^{MNPQR}\}\\
 &  & +\frac{1}{4i}\int \frac{C_{4}\wedge G_{3}\wedge \bar{G}_{3}}{\tau _{I}}.
\end{eqnarray*}

In the above \( \tau =C_{0}+ie^{-\phi }, \) \( G_{3}=F_{3}-\tau H_{3}, \)
with \( F_{3}=dC_{2} \) and \( H_{3}=dB_{2} \). Also \( \tilde{F}_{5}=F_{5}-\frac{1}{2}C_{2}\wedge H_{3}+\frac{1}{2}B_{2}\wedge F_{3} \)
with the self-duality condition \( \tilde{F}_{5}=*\tilde{F}_{5} \)
being imposed by hand at the level of the equations of motion.

In addition there is the action for the D3 branes and orientifold
3-planes in Einstein frame

\[
S_{loc}=\sum _{i}\left( -\int _{i}d^{4}xT_{3}\sqrt{|g^{(4)}|}+\mu _{3}\int _{i}C_{4}\right) .\]

Here the integrals are taken over the 4D non-compact space at a point
\( i \) in the internal manifold and \( T_{3}=\mu _{3}>0\, (<0) \)
for a D-brane (orientifold plane). 

The other equations (Bianchi identities) are

\begin{eqnarray}
d\tilde{F}_{5} & = & H_{3}\wedge F_{3}-\sum _{i}\mu ^{i}_{3}\delta _{6}^{i}\label{F5} \\
dG_{3} & = & -d\tau \wedge H_{3}\nonumber \label{GB} \\
d\Lambda  & = & 0,\, \, \Lambda \equiv e^{4\omega }*_{6}G_{3}-i\alpha G_{3}.\nonumber \\
*F_{5} & =F_{5}\nonumber 
\end{eqnarray}

The self-duality of the five form is satisfied by the following ansatz,

\begin{equation}
\label{5an}
\tilde{F}_{5}=\frac{1}{4!}(1+*)\sqrt{\bar{g}_{4}(x)}d\alpha (x,y)\wedge dx^{0}\wedge ...\wedge dx^{3}
\end{equation}

where \( \alpha (x,y) \) is a scalar function. In addition the three
form fluxes are taken to be non-zero only in the internal directions
and are independent of \( x \) . i.e

\begin{equation}
\label{3flux}
G_{mnp}\ne 0,\, \, \, G_{\mu NP}=0\, \, \, \partial _{x}G=0.
\end{equation}

First we will use the static ansatz (\ref{metric2}) for the metric.
Then the external and (the trace of) the internal components of (\ref{E})
give the following:

\begin{eqnarray}
\bar{R}_{\mu \nu }-e^{4\omega (y)}\tilde{\nabla }^{2}\omega \bar{g}_{\mu \nu }(x) & = & -g_{\mu \nu }(x)\left[ \frac{G_{mnp}\bar{G}^{mnp}}{48\tau _{I}}+\frac{e^{-8\omega }}{4}\partial _{p}\alpha \partial ^{p}\alpha +T_{loc}\right] \label{ext} \\
R^{(6)} & =e^{2\omega }\tilde{R}^{(6)} & -8\partial _{m}\omega \partial ^{m}\omega +6e^{2\omega }\tilde{\nabla }^{2}\omega \label{int} \\
 & = & \frac{G_{mpq}\bar{G}^{mpq}}{8\tau _{I}}+e^{-8\omega }\partial _{m}\alpha \partial ^{m}\alpha +6T_{loc}\nonumber 
\end{eqnarray}

In the above we have defined \( T_{loc}=\frac{1}{4}\sum _{i}T_{3}\frac{\delta ^{6}(y-y^{i})}{\sqrt{g_{6}}} \).
As in the discussion leading to (\ref{effpot}) we can multiply (\ref{ext})
by \( e^{-4\omega } \) and integrate over the internal space to identify
the potential. This tells us that both flux terms give negative definite
contributions to the potential while a positive term can only come
from negative tension terms.

Using \ref{5an}) and (\ref{3flux}) in the Bianchi identity (\ref{F5}),
we also have \begin{equation}
\label{Bi}
\tilde{\nabla }^{2}\alpha =\frac{i}{12\tau _{I}}e^{2\omega }G_{mnp}*_{6}\bar{G}^{mnp}+2e^{-6\omega }\partial _{m}\alpha \partial ^{m}e^{4\omega }+e^{2\omega }\sum _{i}\mu _{3}\frac{\delta ^{(6)}(y-y^{i})}{\sqrt{g^{(6)}}}.
\end{equation}

Using this to eliminate the local term from (\ref{ext}) after using
\( T_{3}=\mu _{3} \) , we have

\begin{equation}
\label{Rbar}
-\bar{R}^{(4)}+\tilde{\nabla }^{2}(e^{4\omega }-\alpha )=\frac{e^{2\omega }}{24\tau _{I}}|iG_{3}-*_{6}G_{3}|^{2}+e^{-6\omega }|\partial (e^{4\omega }-\alpha )|^{2}.
\end{equation}
 This equation is essentially equation (2.30) of GKP except that we
have not set the 4D curvature to zero. Integrating this over the compact
manifold now tells us that irrespective of the existence of negative
tension the 4D curvature is negative definite, implying that the potential
is negative definite except at its extremum (maximum) where it is
zero. The problem again is that the ansatz (\ref{metric2}) is too
restrictive. While it can be used to draw conclusions about the static
solutions (in particular that to get flat space in four dimensions
and to have a Ricci flat tilde metric in six one needs the net tension
of the local sources to be negative) it cannot be used to calculate
the potential unless the potential for the volume modulus has a critical
point.

Let us now re-derive these equations by using the more general ansatz
(\ref{metric3}) so that the time dependence of the moduli is explicit.
First we note that our ansatz for the fluxes gives \( T_{\mu n}=0 \)
so the corresponding Einstein equation is \( R_{\mu n}=0 \). We now
write\[
\tilde{g}_{mn}(x,y)=\tilde{g}_{mn}(y)+z_{i}(x)\phi ^{i}_{mn}(y)+...\]
 where the \( \phi _{mn}^{i} \) are zero modes of the six dimensional
Laplacian and the ellipses represent non-zero modes. Thus the \( z_{i} \)
are all other moduli apart from the volume modulus and the requirement
that the volume modulus has been factored out implies that to linearized
order in the \( z_{i} \) gives the \( \phi 's \) should be traceless.
Then from (\ref{Rmun}) we have\begin{equation}
\label{cons1}
\partial _{p}\omega (y)(2\partial _{\mu }u(x)\delta ^{p}_{n}+\partial _{\mu }z_{i}(x)\tilde{g}^{rp}\phi ^{i}_{rn})+...=0
\end{equation}
 where the ellipses represent higher KK modes which have been set
to zero in our current ansatz.

The Bianchi identity now takes the form \begin{equation}
\label{Bi2}
\tilde{\nabla }^{2}\alpha =\frac{i}{12\tau _{I}}e^{8\omega -4u}\widetilde{G_{mnp}*_{6}\bar{G}^{mnp}}+8\widetilde{\partial _{m}\alpha \partial ^{m}\omega }+e^{8\omega -4u}\sum _{i}\mu _{3}\frac{\delta ^{(6)}(y-y^{i})}{\sqrt{\widetilde{g^{(6)}}}}
\end{equation}

In addition there is also a constraint similar to (\ref{cons1})

\begin{equation}
\label{cons2}
(d_{x}u+\tilde{g}^{mn}d_{x}\tilde{g}_{mn})d_{y}\alpha +d_{x}d_{y}\alpha =0
\end{equation}
Inserting (\ref{Rmunu})(\ref{Rmn}) with \( \tilde{R}^{(6)}=0 \),
in (\ref{E1}) and using (\ref{Bi2}) to eliminate the local source
term in \( T_{\mu \nu } \), we get

\begin{eqnarray}
\tilde{R}_{\mu \nu }-\frac{1}{2}\tilde{R}^{(4)}\tilde{g}_{\mu \nu } & =-\frac{1}{4}\tilde{g}_{\mu \nu }[\frac{e^{2\omega }}{12\tau _{I}}|iG_{3}-*_{6}G_{3}|^{2}+e^{-4\omega -8u}\widetilde{\partial _{m}(\alpha -e^{4\omega })^{2}} & \label{R41} \\
 & +e^{-8u}(\widetilde{\nabla ^{2}}(\alpha -e^{4\omega })+e^{-4\omega }\partial _{m}e^{4\omega }\partial ^{m}(\alpha -e^{4\omega }))]+...
\end{eqnarray}

Also as before, the ellipses represent terms which are first order
in 4-space-time and we stress that there are no second order in time
derivatives of the moduli in this equation. We observe now that the
LHS of this equation is \( y \) independent so that the same must
be true of the RHS. Thus the measure that we choose to integrate this
equation over the internal manifold is ambiguous up to purely \( y \)
dependent factors. However by substituting the metric ansatz (\ref{metric3})
into the ten dimensional action it would appear that the natural measure
is \( e^{-4\omega }\sqrt{\widetilde{g^{(6)}}} \) . The above equation
then takes the form 

\[
\tilde{R}_{\mu \nu }-\frac{1}{2}\tilde{R}^{(4)}\tilde{g}_{\mu \nu }=8\pi G_{N}T_{\mu \nu }^{(4)}\]
where \( 1/16\pi G_{N}=\int e^{-4\omega }\sqrt{\tilde{g}^{6)}} \)
and \( T_{\mu \nu }^{(4)}=-V\tilde{g}_{\mu \nu }+... \) where the
four dimensional potential may be identified as\[
V=\int d^{6}y\sqrt{\tilde{g}^{(6)}}\frac{1}{2}e^{-4\omega }\left[ \frac{e^{8\omega -12u}}{12\tau _{I}}\widetilde{|iG_{3}-*_{6}G_{3}|^{2}}+e^{-4\omega -8u}\widetilde{\partial _{p}(\alpha -e^{4\omega })^{2}}\right] \]
 This is clearly positive definite and indeed it is tempting to fix
the warp factor by setting \( e^{4\omega }=\alpha +const. \) (this
relation was imposed \emph{ab initio} in \cite{DeWolfe:2002nn} in
their otherwise similar derivation of this action) thus giving us
the desired potential of \cite{Taylor:1999ii}\cite{Giddings:2001yu}.
Let us now count the number of equations and the number of 6 dimensional
fields that we have in the above equation. The two equations after
(\ref{F5}) determine the (complex) flux \( G_{3} \) . Given the
sources (whose total charge is determined in terms of the fluxes by
the integrated Bianchi identity) the local form (\ref{Bi}) maybe
thought of as determining \( \alpha  \) . The above relation then
fixes \( \omega  \). We will now see that it is in fact required
for consistency.

The effective four dimensional action that may be deduced from the
above is\[
S=\int d^{4}x\sqrt{\tilde{g}^{(4)}[\tilde{R}^{4}}-24\tilde{g}^{\mu \nu }\partial _{\mu }u\partial _{\nu }u-\frac{1}{4}\tilde{g}^{\mu \nu }\partial _{\mu }z^{i}\partial _{\nu }z^{i}-V]\]
 The equation of motion for the modulus \( u \) is then\begin{eqnarray}
6\tilde{\nabla }^{2}_{x}u(x) & =\partial _{u}V & \nonumber \\
 & =\int d^{6}y\sqrt{\tilde{g}^{(6)}} & e^{-4\omega }\left[ -\frac{e^{8\omega -12u}}{2\tau _{I}}\widetilde{|iG_{3}-*_{6}G_{3}|^{2}}-4e^{-4\omega -8u}\widetilde{(\partial _{p}(\alpha -e^{4\omega }))^{2}}\right] \label{ueom1} 
\end{eqnarray}

This equation should be consistent with that directly obtained from
the ten dimensional equation by using (\ref{Rmn}) in the trace over
the internal space of (\ref{E}) putting \( \tilde{R}^{(6)}=0 \)
and using the Bianchi identity (\ref{Bi2}) with \( T_{3}=\mu _{3} \)
. Thus we get,\begin{equation}
\label{R6B}
-6\tilde{\nabla }^{2}_{x}u(x)=\frac{e^{8\omega -12u}}{16\tau _{I}}\widetilde{|iG_{3}-*_{6}G_{3}|}^{2}+e^{-4\omega -8u}[\widetilde{(\partial _{m}\alpha )}^{2}+2\widetilde{(\partial _{m}e^{4\omega })}^{2}-3\widetilde{\partial _{m}\alpha \partial ^{m}}e^{4\omega }]-\frac{3}{2}e^{-8u}\tilde{\nabla }_{y}^{2}(e^{4\omega }-\alpha ).
\end{equation}

integrating over the internal space with the same measure as before
we have a discrepancy with (\ref{ueom1}) unless \( e^{4\omega }=\alpha +const \)
. In this case \begin{equation}
\label{potential}
V=\int d^{6}y\sqrt{\tilde{g}^{(6)}}\frac{e^{4\omega -12u}}{24\tau _{I}}\widetilde{|iG_{3}-*_{6}G_{3}|^{2}}
\end{equation}

\begin{equation}
\label{ueqn}
-6\tilde{\nabla }^{2}_{x}u(x)=\frac{e^{8\omega -12u}}{16\tau _{I}}\widetilde{|iG_{3}-*_{6}G_{3}|}^{2},
\end{equation}

and there is no critical point for the potential for the volume modulus,
which is then runaway as long as the flux is not imaginary self dual
(ISD) (i.e. does not satisfy \( iG_{3}=*_{6}G_{3} \)). As shown in
\cite{Taylor:1999ii}\cite{Giddings:2001yu} the above potential can
be written in the four dimensional \( \cal N \) =1 supergravity form

with a superpotential \( W=-\int \Omega \wedge G_{3} \) and a Kahler
potential \( K=-\ln [-i(\tau -\bar{\tau })]-3\ln [-i(\rho -\bar{\rho })]-\ln \left( -i\int \Omega \wedge \bar{\Omega }\right)  \)
where \( \Omega  \) is the holomorphic three form on the Calabi-Yau
space, and where \( D_{i} \) is the Kahler covariant derivative and
\( K^{i\bar{j}} \) is the inverse Kahler metric. 

This then is the resolution of the puzzle that we uncovered regarding
the derivation of the potential. However equation (\ref{ueqn}) still
highlights a problem. The point is that the LHS is independent of
\( y \) and therefore also the RHS. This seems rather restrictive.
In fact we've already seen a related problem. It comes from the constraints
(\ref{cons1})(\ref{cons2} ) and the above relation between \( \alpha  \)
and \( \omega  \), which tell us that away from the ISD point (where
the moduli will have space-time dependence) the only solution for
the warp factor is the trivial one i.e \( \partial _{m}\omega =0 \).
This implies that away from the extremum of the potential we cannot
consistently set the KK modes to zero unless the warp factor is trivial. 

To see this in more detail let us choose a more general metric ansatz
to replace (\ref{metric3}). As a first attempt let us just replace
\( u(x) \) by \( u(x,y) \) and also keep the general \( (x,y) \)
dependence of the metric \( \tilde{g}_{mn} \) . Then the restriction
to constant \( \omega  \), coming from (\ref{cons1})(\ref{cons2})
will no longer apply. We can now work out again the equations of motion
using (\ref{Rmunu})(\ref{Rmn}). The ansatz for the five-form is
modified as follows.

\[
\tilde{F}_{5}=\frac{1}{4!}(1+*)\sqrt{\tilde{g}_{4}(x)}d\alpha (x,y)\wedge dx^{0}\wedge ...\wedge dx^{3}\]

This amounts to the replacement of \( \alpha  \) by \( e^{-12u(x,y)}\alpha  \)
and of course leads to the same equations as before for \( u \) independent
of \( y \) . In the more general case the Bianchi identity (\ref{Bi2})
is replaced by\begin{equation}
\label{Bi3}
\tilde{\nabla }^{2}\alpha =\frac{i}{12\tau _{I}}e^{8\omega -16u}\widetilde{G_{mnp}*_{6}\bar{G}^{mnp}}+\tilde{g}^{mn}(8\partial _{m}\omega -16\partial _{m}u)\partial _{n}\alpha +e^{8\omega -16u}\sum _{i}\mu _{3}\frac{\delta ^{(6)}(y-y^{i})}{\sqrt{\widetilde{g^{(6)}}}}
\end{equation}

and we also have again (\ref{cons2}).

After a straightforward though tedious calculation the effective action
including these KK modes can be obtained after using the above to
eliminate the local source term and imposing  

\begin{equation}
\label{alphom}
\partial _{m}\alpha =\partial _{m}e^{4\omega (y)-12u(x,y)}
\end{equation}

Note that this is the same as the previous constraint except since
we have redefined \( a \). 

The action then becomes\begin{eqnarray*}
\frac{1}{16\pi G_{N}}\int \sqrt{\tilde{g}^{(4)}}\tilde{R}^{(4)}+\int d^{4}x\sqrt{\tilde{g}^{(4)}}\int d^{6}y\sqrt{\tilde{g}^{(6)}}e^{-4\omega (y)} & [-24\tilde{g}^{\mu \nu }\partial _{\mu }u(x,y)\partial _{\nu }u(x,y) & \\
-\frac{1}{4}\tilde{g}^{pm}\tilde{g}^{rn}\tilde{g}^{\mu \nu }\partial _{\mu }\tilde{g}_{mr}(x,y)\partial _{\nu }\tilde{g}_{np}(x,y) & -\int d^{4}x\sqrt{\tilde{g}^{(4)}}V[u,\tilde{g}_{mn}] & 
\end{eqnarray*}

with \begin{equation}
\label{pot2}
V[u,\tilde{g}_{mn}]=\int \sqrt{\tilde{g}^{(6)}}d^{6}y\left[ \frac{e^{4\omega (y)-12u(x,y)}}{24\tau _{I}}\widetilde{|iG_{3}-*_{6}G_{3}|^{2}}-80e^{-8u}\widetilde{(\partial _{m}u(x,y))^{2}}\right] 
\end{equation}

The question then is the interpretation of the KK excitations of \( u(x,y) \)
since they appear to give a dangerous negative definite contribution
to the potential. In fact observe that now there is a critical point
for the volume modulus and its KK excitations) and they can be integrated
out and we would again get the negative potential puzzle that we pointed
out after (\ref{Rbar}). 

To clarify what goes wrong let us consider Kaluza-Klein compactification
in the absence of fluxes. Let us take the metric ansatz:\begin{equation}
\label{metric4}
ds^{2}=e^{-6u(x,y)}\tilde{g}_{\mu \nu }(x)dx^{\mu }dx^{\nu }+e^{2u(x,y)}\tilde{g}_{mn}(y)dy^{m}dy^{n}
\end{equation}

The internal metric \( \tilde{g}_{mn} \) is defined by requiring
\( \partial _{\mu }\det [\tilde{g}_{mn}]=0 \) and the prefactor in
the four dimensional metric is put there as before to eliminate mixing
between the dimensional graviton (and its KK tower) and the conformal
mode of the internal space \( u(x,y) \) i.e. \( \sqrt{g}g^{\mu \nu }=\sqrt{\tilde{g}^{(4)}}\sqrt{\tilde{g}^{(6)}}\tilde{g}^{\mu \nu } \)
. To compute the effective action for \( u \) we may ignore the \( x \)
dependence of \( u \) and \( \tilde{g}_{mn} \) and use standard
formulas for conformal transformations. (The calculation may be done
without this assumption by using (\ref{Rmunu})(\ref{Rmn}) but is
then much more tedious). We then have the action

\begin{equation}
\label{puzzle}
\int \sqrt{g}R=\int \sqrt{\tilde{g}^{(4)}(x)}\sqrt{\tilde{g}^{(6)}(y)}[\tilde{R}(x)^{(4)}-24\tilde{g}^{\mu \nu }\partial _{\mu }u(x,y)\partial _{\nu }u(x,y)+e^{-8u}(\tilde{R}^{(6)}+8\tilde{g}^{mn}\partial _{m}u\partial _{n}u)]
\end{equation}

The problem is that this appears to give a negative potential for
the conformal factor \( u \) independently of the sign of the curvature
since one can go to a high enough Kaluza-Klein excitation to make
the second term dominate over the first. In fact this effect will
persist for all physically interesting compactifications and appears
to imply that all such compactifications are unstable. To resolve
this let us analyze the linearized equations around a flat background
say \( R^{4}\times T^{6} \) . (This line of investigation was suggested
to the author by Gary Horowitz). The equations of motion for pure
(10 dimensional) gravity are at linearized order, 

\[
-\frac{1}{2}\nabla ^{2}h_{MN}+\frac{1}{2}(\partial _{M}\partial ^{L}h_{NL}+\partial _{N}\partial ^{L}h_{ML}-\partial _{M}\partial _{N}h^{L}_{\, L})=0,\]
 where we have written the \( D \) dimensional metric as \( g_{MN}=\eta _{MN}+h_{MN}. \)
The equations are clearly gauge invariant under the transformations

\[
h_{MN}\rightarrow h_{MN}+\partial _{M}\xi _{N}+\partial _{M}\xi _{N}.\]

The point is that even though the compactification breaks the original
diffeomorphism invariance to that appropriate to the product space,
the transformations induced on the fields by the original ten dimensional
diffeomorphims still remain invariances of the action. Now starting
from some arbitrary configuration \( h_{MN} \) we can always go to
a gauge (the so-called harmonic gauge) in which \( \partial ^{L}h_{ML}-\frac{1}{2}\partial _{M}h^{L}_{\, L}=0 \).
(This is done by choosing \( \xi _{M}=\nabla ^{-2}(\partial ^{L}h_{ML}-\frac{1}{2}\partial _{M}h^{L}_{\, L}) \)
in terms of the original configuration). In this gauge the equations
of motion reduce to \( \nabla ^{2}h_{MN}=\nabla ^{2}_{x}h_{MN}+\nabla ^{2}_{y}h_{MN}=0 \).
Expanding in modes on the internal space \( \nabla ^{2}_{y}\rightarrow -m_{r}^{2} \)
and we have \( \nabla ^{2}_{x}h^{(r)}_{MN}-m_{r}^{2}h^{(r)}_{MN}=0 \)
, which means that all the KK modes of the gravity sector are non-tachyonic.
On the other hand the linearized equations coming from (\ref{puzzle})
are (after writing \( \tilde{g}_{\mu \nu }(x)=\eta _{\mu \nu }+\tilde{h}_{\mu \nu }(x) \)
,\begin{eqnarray*}
-\frac{1}{2}\nabla _{x}^{2}\tilde{h}_{\mu \nu }+\frac{1}{2}(\partial _{\mu }\partial ^{\lambda }\tilde{h}_{\nu \lambda }+\partial _{\nu }\partial ^{\lambda }\tilde{h}_{\mu \lambda }-\partial _{\mu }\partial _{\nu }\tilde{h}^{\lambda }_{\, \lambda }) & = & 0\\
-3\nabla ^{2}_{x}u(x,y)+\nabla ^{2}_{y}u(x,y) & = & 0
\end{eqnarray*}

and the second equation appears to tell us that there is a whole KK
tower of tachyons. However instead of deriving the equations from
the action (\ref{puzzle}) let us substitute the metric ansatz (\ref{metric4})
into the ten-dimensional equations of motion. From the equation \( R_{\mu \nu }-\frac{1}{2}Rg_{\mu \nu }=0, \)
(i.e. the four-space projection of the ten-dimensional equation) we
have, 

\[
-\frac{1}{2}\nabla _{x}^{2}\tilde{h}_{\mu \nu }(x)+\frac{1}{2}(\partial _{\mu }\partial ^{\lambda }\tilde{h}_{\nu \lambda }(x)+\partial _{\nu }\partial ^{\lambda }\tilde{h}_{\mu \lambda }(x)-\partial _{\mu }\partial _{\nu }\tilde{h}^{\lambda }_{\, \lambda }(x))-4\nabla ^{2}_{y}u(x,y)\eta _{\mu \nu }=0\]

What we have from the action is in fact the version of this that is
integrated over the internal manifold. However the above equation
implies, since the ansatz has the 4 dimensional metric \( \tilde{g}_{\mu \nu } \)
independent of \( y \), that all the KK excitations of \( u \) are
all zero! In fact in order to have a consistent set of linearized
equations one needs to keep all the KK excitations. For instance it
is not enough to remedy the situation by changing the ansatz to include
the KK excitations of the four metric (i.e. replace  \( g_{\mu \nu }(x)\rightarrow g_{\mu \nu }(x,y) \)
). One needs to include also the off diagonal terms \( g_{m\nu }(x,y) \).
Otherwise the corresponding equation of motion becomes a constraint
on the other fields. In that case we have the full gauge invariance
of the ten dimensional theory and as argued earlier all the KK modes
will have positive squared masses.

The same then is true for the calculation with the fluxes. It is inconsistent
just to have kept the KK excitations of the volume modulus \( u \).
Indeed the upshot of our investigation is that the when one takes
into account all the ten dimensional equations a non-trivial warp
factor requires one to include all the KK modes. This means in particular
that in addition to the desired term in (\ref{potential}) there are
terms such as the second term in (\ref{pot2}) but with a positive
sign. Thus there will be no critical point for the volume modulus
as in the case where we ignored the KK excitations. The problem however
is that it is not possible to set the KK excitations to zero without
at the same time requiring that the warp factor be trivial. In other
words strictly speaking the derivation of the potential (\ref{potential})
is valid only for a trivial warp factor while if one insists that
it be non-trivial (as would be the case in the presence of local sources)
one needs to keep also the infinite towers of KK excitations. When
one integrates out the KK modes, for consistency one would have to
integrate out the dilaton and the complex structure moduli as well,
and one is then constrained to be at the point where the potential
is zero.

\section{Conclusions}

Our analysis shows that for a non-trivial warp factor there is no
meaningful way of deriving a potential just for the moduli. The consistent
reduction to four dimensional massless modes leads (ignoring the gauge
sector) to a theory of four dimensional supergravity coupled to the
volume modulus with no potential, 

\begin{equation}
\label{4daction}
S=\int d^{4}x\sqrt{g}\left[ R-3\frac{\partial _{\mu }\rho \partial _{\mu }\rho }{\rho -\bar{\rho }}\right] 
\end{equation}

where \( \rho =a+ie^{4u(x)} \). This is of course a perfectly acceptable
supergravity and the corresponding superpotential for \( \rho  \)
can in fact be a constant since given the form of the Kahler potential
\begin{equation}
\label{K}
K(\rho \bar{,}\bar{\rho })=-3\ln [-i(\rho -\bar{\rho })]
\end{equation}

a vanishing potential is consistent with a constant superpotential.
The latter would be certainly given by the expression \( W=\int G_{3}\wedge \Omega  \)
at the minimum (see discussion after (\ref{potential})) when the
warp factor is trivial. On the other hand for non-trivial warp factor,
one cannot set the KK modes to zero and it is not immediately clear
that one can obtain this expression. There is a general argument in
\cite{Gukov:1999ya} that uses the fact that on crossing a BPS domain
wall, the superpotential changes by the tension of the wall, to show
that this change is given by the change in the above expression for
\( W \) . But this still leaves an undetermined arbitrary constant.
(The fact that the real part of \( \rho  \) is an axion which has
a shift symmetry implies that \( W \) must be independent of \( \rho  \)
to all orders in perturbation theory). However as pointed out to the
author by S. Kachru, if such a constant is present, it is there even
in the large radius limit, and would break supersymmetry in violation
of the conditions for supersymmetry in the ten dimensional theory
given in \cite{Becker:1996gj}\cite{Dasgupta:1999ss}\cite{Greene:2000gh}\cite{Grana:2000jj}.
Thus we would have to conclude that the expression for the superpotential
at the minimum is unchanged even in the case of non-trivial warping.

The upshot is that after integrating out the the KK modes and the
complex structure moduli one is left with (\( \cal N \) =1) supegravity
coupled to one modulus field namely the size of the internal space,
and a non-zero superpotential whose natural value is \( O(1) \) in
string units. These moduli and the dilaton are determined by the fluxes
that are turned on. They are of course quantized and therefore the
choices are discrete but nevertheless there is a very large (perhaps
infinite) number of choices, since there is no physical principle
that tells us that one flux configuration is preferable to another.
Of course the main achievement of these flux compactifications is
that there are no massless fields corresponding to the complex structure
moduli. In this respect compactifications with fluxes are similar
to non-geometric compactifications such as asymmetric orbifolds. In
these there are no geometric moduli but the dilaton is still a massless
field.

The fixing of the volume modulus (or in the case of non-geometric
compactifications - the dilaton) and the breaking of supersymmetry
can now be done in the standard way - namely by gaugino condensation
in the gauge theory on the branes. Thus with multiple condensates
one has the so-called racetrack models \cite{Krasnikov:1987jj} with
superpotentials which are exponentials of the (field dependent) effective
gauge coupling \( f=\tau _{0}+\alpha \rho  \) where \( \tau _{0} \)
is the (fixed) value of the dilaton and the \( \rho  \) depencence
comes from threshold effects with \( \alpha  \) a model dependent
constant (we ignore the dependence on the other moduli since they
are also fixed like the dilaton) . Together with the arbitrary constant
mentioned above, one would then have a superpotential of the form

\[
W=W_{0}+\sum _{i}A_{i}e^{ib_{i}\rho },\]

where \( b_{i} \) are beta function coefficients and the \( A_{i} \)
are calculable prefactors of the corresponding instanton calculation.
Actually such non-perturbatively generated superpotentials should
in some sense be dual to the ones generated by the fluxes as in the
non-compact case discussed by Dijkgraaf and Vafa (work in progress
with R. Brustein and E. Novak). Thus it should be possible to generate
the entire superpotential above from one point of view or the other.
Since so far we do not have a complete understanding of this we will
content ourselves with the above prescription.

With more than two condensates one should in principle be able to
get small SUSY breaking and a small cosmological constant, though
a concrete example has not yet been found. An alternative method of
stabilization of the volume modulus is the so-called Kahler stabilization,
where one may use the quantum corrections to K to get a critical point
even with one condensate and no constant \cite{Banks:1994sg}. However
in the absence of an adjustable constant it is very difficult to see
how one can get a small positive cosmological constant in either mechanism.

The discussion above is similar in spirit to that in \cite{Kachru:2003aw},
but is different in one respect. The mechanism for the stabilization
and SUSY breaking that we propose is the standard one used many times
in the literature to stabilize the dilaton, except that here it is
used to stabilize the volume modulus, whereas in \cite{Kachru:2003aw}
SUSY is broken by including a \( \bar{D} \) brane which breaks SUSY
explicitly. Of course in either case  to get a phenomenologically
viable result one needs to fine tune \( W_{0} \) and \( \textrm{A}_{i} \),
and since these are in principle calculable the question of whether
such a result can actually be obtained is still open. Finally we observe
that from a cosmological point of view the problem with the steep
potentials that come from the above superpotential (with small or
zero \( W_{0} \)) is not so much the possibility of quantum tunneling
from a deS minimum but the classical overshoot problem of \cite{Brustein:1993nk}.

In summary our conclusions are as follows. 

\begin{itemize}
\item Compactification with fluxes of type IIB string theory on a Calabi-Yau
orientifold leads to a four dimensional theory where all complex structure
moduli except the volume modulus is fixed, with the possibility of
the dilaton being fixed at some weak coupling value. The conflict
between the no-go theorem and the existence of a positive potential
gets resolved, since in these theories the volume modulus is not stabilized
classically and has a runaway potential as long as the other moduli
are not at the minimum of the potential.
\item When the warp factor is trivial there is an unambiguous derivation
of the potential for the complex structure moduli. However this case
would be relevant only if one can assume that the D-branes are in
some sense smeared over the Calbi-Yau space. One might imagine this
to be the case if \( \rho  \) is such that the compactification scale
is only slightly greater than the string scale. In general the warp
factor will be non-trivial and then there is no clear derivation of
the potential, since the KK modes cannot be set to zero without making
the moduli space time independent. In other words the four dimensional
theory would contain the infinite set of KK modes as well and setting
them to zero would require the moduli to be fixed at values where
the potential is zero. This could then just as well have been derived
directly from the observation that the condition for getting a Killing
spinor in the presence of fluxes implies that the complex structure
moduli and the dilaton have fixed values \cite{Becker:1996gj}\cite{Dasgupta:1999ss}\cite{Greene:2000gh}\cite{Grana:2000jj}\cite{Frey:2002hf}.
\item What one has in four dimensions is supergravity coupled to the (complex)
volume modulus field \( \rho  \) with zero potential and supersymmetry
broken by a constant superpotential. One can then invoke non-perturbative
effects (namely gaugino condensation) to get a potential for this
modulus.
\item The potentials that one gets in such models suffer from the overshoot
problem of \cite{Brustein:1993nk}. In other words for this modulus
to settle down in the classically stable local minimum such as the
one discussed in \cite{Kachru:2003aw}, requires fine tuning of initial
conditions. The only way that we know of avoiding this is a situation
in which when the infinite set of non-perturbative corrections are
properly taken into account, the moduli are stabilized in the region
where the sizes of extra dimensions are fixed at the string scale.
This possibility and the corresponding cosmology are discussed in
\cite{Brustein:2000mq},\cite{Brustein:2002mp}.
\end{itemize}

\section{Acknowledgments}

I wish to thank Ramy Brustein, Michael Duff, Alex Flournoy, Shamit
Kachru, Juan Maldacena, Eric Novak, Kostas Skenderis, and especially
Steve Giddings, Gary Horowitz and Ashoke Sen, for useful discussions.
I also wish to thank the Director of the Newton Institute Cambridge
and the organizers of the workshop on M theory where this work was
begun. This research is supported in part by the United States Department
of Energy under grant DE-FG02-91-ER-40672.

\bibliographystyle{apsrev}
\bibliography{myrefs}

\end{document}